# Electrostatic Interaction in Stochastic Electrodynamics

Ion Simaciu[1,a], Zoltan Borsos[1,b], Viorel Drafta[2] and Gheorghe Dumitrescu[3]

[1] Retired lecturer, Petroleum-Gas University of Ploiești, Ploiești 100680, Romania

[2] Independent researcher

[3] High School Toma N. Socolescu, Ploiești, Romania

E-mail: [a] isimaciu@yahoo.com; [b] borzolh@upg-ploiesti.ro

**Abstract**
Assuming the charged particle to be a two-dimensional oscillator that scatters the classical background of zero-point field one can deduce the Coulomb force of the two interacting particles. The correct deduction of the force is conditioned by the equality between the natural angular frequency of the oscillator and the angular frequency of Zitterbewegung.

1. **Introduction**

The Stochastic Electrodynamics (SED) studies the interaction of charged particles with the subquantum environment. In SED, the subquantum environment is supposed to be the background of electromagnetic waves with random phases. This background is the analogous of the background of quantum fluctuations with zero temperature (or Zero point field (ZPF) in quantum electrodynamics (QED).

SED was founded by Brafford, Surdin, Tzara, Taroni [1- 3] as a consequence of the Wheeler - Feynman theory of radiation [4]. The theory was later developed by T. Marshall [5, 6], TH Boyer [7- 9] and L. de la Peña-Auerbach and A. M. Cetto [10, 11].

According to the analysis made by T. H. Boyer [12, 13] on SED, it is based on the principles of Classical Electrodynamics (CED) but with different boundary conditions. Although it does not use the concepts of quantum and quantification but reaches the same results obtained by QED: the van der Waals forces and Casimir effect [14-18], the Planck blackbody spectrum [19-21], the third law of thermodynamics [22], rotator and oscillator specific heats [23], the diamagnetism of free particles [24, 25], the thermal effect of acceleration [26-28], the stationary state of microscopic systems [29-31], the Lamb shift (radiative corrections) [32], etc.

This paper generalizes the particle model, in SED picture, in order to study the electrostatic interaction of two different particles. According to the model developed in [33] and improved in this paper, the electrostatic interaction between two charged particles is figured as an effect of interaction between the CZPF background and charged particles.

Since the concept of scattering cross-section must be used in this paper, then the intensity of the scattered radiation background is averaged in time and does not depend on the oscillation phases of the two oscillators [34]. For this reason, the model does not allow the explanation that there are two types of electric charge.

Both attractive and repulsive behaviour of the forces may occur between oscillators if we consider that the two oscillators can oscillate in phase or in phase opposition. The phenomenon is analogous to the forces of the interaction between two oscillating bubbles, i.e. the secondary Bjerknes forces [35-37].

At the microscopic level, the electron model is characterized by a scattering cross-section dependent on angular frequency $\omega$ [33]. If the electron is at rest, the averaged electromagnetic scattering cross-section (Appendix 3) is proportional to the square of the electrostatic radius, i.e., the Thomson cross-section [38].

In the elementary charged particle model in Stochastic Electrodynamics [33], the particle at rest is an oscillator that scatters the CZPF background and the scattering cross-section depends on the angular frequency $\omega$ of the scattered electromagnetic wave and the natural angular frequency of the oscillator $\omega_i$. In this model, electric charge is a measure of the ability of the microscopic system to scatter the stochastic background of radiation. This interpretation of electric charge is supported by the connection between discrete electric charge, relativity and radiation thermodynamics, highlighted, in the classical (non quantum) context, in the paper [39].

## 2. Electrostatic force in SED

We will further model a charged particle (in particular, the electron) as a two-dimensional oscillator that scatters CZPF (Classical Zero Point Field) background radiation. Our derivation is related to the model of the non-relativistic charged oscillator [38- Sch.17.8, pp. 801-806].

The oscillator has the mass $m_i$ and the electric charge $q_i = n_i |q_e|$, ($n_i = 1/3, 2/3, 1$). Its scattering cross-section with the plane electromagnetic wave [33, 38] is

$$\sigma_i(\omega) = \frac{6\pi c^2 \omega^4 \Gamma_i^2}{\omega_i^4 \left[ \left( \omega_i^2 - \omega^2 \right)^2 + \left( \Gamma_i^2 \omega^6 \right) / \omega_i^4 \right]}, \quad i = 1, 2. \quad (1)$$

In previous relationships, $\Gamma_i$ is the radiative decay constant

$$\Gamma_i = \frac{2R_i}{3c} \omega_i^2, \quad (2)$$

and the total decay constant $\Gamma_{ti}(\omega) = \Gamma_i' + (\omega^2/\omega_i^2)\Gamma_i \cong \Gamma_i$ because the absorptive width (absorptive decay constant) $\Gamma_i' \ll \Gamma_i$. In Eq. (2), $R_i$ is the electrostatic radius of the particle,

$$R_i = \frac{q_i^2}{4\pi\varepsilon_0 m_i c^2} = \frac{n_i^2 e^2}{m_i c^2}. \quad (3)$$

Replacing $\Gamma_i$ from Eq. (2) in Eq. (1), yields

$$\sigma_i(\omega) = \frac{8\pi R_i^2 \omega^4}{3\left[ \left( \omega_i^2 - \omega^2 \right)^2 + \left( 2R_i / (3c) \right)^2 \omega^6 \right]}, \quad i = 1, 2. \quad (4)$$

According to Eq. (26) of paper [33], the elementary force of interaction between two oscillators that scatter the CZPF background has the expression

$$\delta F_{12}(r, \omega) = \frac{\rho(\omega) \sigma_1(\omega) \sigma_2(\omega) d\omega}{3\pi r^2}. \quad (5)$$

The CZPF background is characterized by the spectral density of the energy [8]

$$\rho(\omega, T = 0) = \rho(\omega) = \frac{\hbar \omega^3}{2\pi^2 c^3}. \tag{6}$$

Replacing Eq. (4) and Eq. (6) in Eq. (5) and integrating, one obtaines

$$F_{12}(r) = \frac{2^5 \hbar R_1^2 R_2^2}{3^3 \pi r^2 c^3} \int_0^\Omega \frac{\omega^{11} d\omega}{\left[(\omega_1^2 - \omega^2)^2 + (2R_1/(3c))^2 \omega^6\right]\left[(\omega_2^2 - \omega^2)^2 + (2R_2/(3c))^2 \omega^6\right]} = \frac{2^5 \hbar R_1^2 R_2^2}{3^3 \pi r^2 c^3} I_\Omega, \Omega > \omega_2 \neq \omega_1. \tag{7}$$

The integral $I_\Omega$ in Eq. (7) is estimated according to the saddle point method [40 - Sch. 41.2, 33] that allows an analytical evaluation. We consider that the particles have the angular frequency approximately equal. This is the situation for the constituent particles of nucleons, the quarks [42]. Under these conditions, $\omega_2 \cong \omega_1 = \omega_0$, the integral $I_\Omega$, according to Appendix 1, is

$$I_\Omega = \int_0^\Omega \frac{\omega^{11} d\omega}{\left[(\omega_0^2 - \omega^2)^2 + (2R_1/(3c))^2 \omega^6\right]\left[(\omega_0^2 - \omega^2)^2 + (2R_2/(3c))^2 \omega^6\right]} \cong \frac{3^3 \pi c^3 \omega_0}{2^4 R_1 R_2 (R_1 + R_2)}, \Omega > \omega_0. \tag{8}$$

Replacing Eq. (8) in Eq. (7), yields

$$F_{12}(r) = \frac{2^5 \hbar R_1^2 R_2^2}{3^3 \pi r^2 c^3} \frac{3^3 \pi c^3 \omega_0}{2^4 R_1 R_2 (R_1 + R_2)} = \frac{2\hbar \omega_0 R_1 R_2}{r^2 (R_1 + R_2)}. \tag{9}$$

Replacing Eq. (3) in Eq. (9), it leads to

$$F_{12}(r) = \frac{2^5 \hbar R_1^2 R_2^2}{3^3 \pi r^2 c^3} \frac{3^3 \pi c^3 \omega_0}{2^4 R_1 R_2 (R_1 + R_2)} = \left(\frac{n_1 n_2 e^2}{r^2}\right) \frac{2\hbar \omega_0 n_1 n_2}{c^2 (n_1^2 m_2 + n_2^2 m_1)}. \tag{10}$$

The expression for the force (10) is like the expression for the Coulomb force, $F_{12}(r) = n_1 n_2 e^2 / r^2$ if

$$\frac{2\hbar \omega_0 n_1 n_2}{c^2 (n_1^2 m_2 + n_2^2 m_1)} = 1 \text{ or } \hbar \omega_0 = \frac{c^2 (n_1^2 m_2 + n_2^2 m_1)}{2 n_1 n_2} = \frac{c^2}{2}\left(\frac{n_2}{n_1} m_1 + \frac{n_1}{n_2} m_2\right). \tag{11}$$

The quarks are characterized by two types of mass [42, 43]: the current/naked/bare quark mass ($m_{un} = 1.8 - 2.8 \text{ MeV}/c^2 = 3.52 - 5.48 m_e$, $n_1 = n_u = 2/3$; $m_{dn} = 4.3 - 5.2 \text{ MeV}/c^2 = 8.41 - 10.17 m_e$, $n_2 = n_d = 1/3$) and the constituent quark mass ($m_{dc} = 340 \text{ MeV}/c^2 = 665.36 m_e$, $m_{uc} = 336 \text{ MeV}/c^2 = 657.53 m_e$). In the case of up and down quarks, the expression (11) is just the resonance condition.

When particles have different natural angular frequencies (the case of interactions between the protons, electrons, muons ($n_p = n_e = n_\mu = 1$), etc.), the integral must be averaged for the two natural angular frequencies $I_\Omega = (I_{\Omega 1} + I_{\Omega 2})/2$. According to the results of Appendix 2

$$I_{\Omega av} = \frac{1}{2}(I_{\Omega 1} + I_{\Omega 2}) = \frac{1}{2}\left(\frac{3\pi c\omega_1^7}{2^2 \omega_2^4} + \frac{3^3 \pi c^3 \omega_2}{2^4 R_1^2 R_2}\right) \cong$$
$$\frac{3^3 \pi c^3 \omega_2}{2^5 R_1^2 R_2}, \Omega > \omega_2 > \omega_1,$$
(12)

Replacing Eq. (12) in Eq. (7), it follows

$$F_{12}(r) = \frac{2^5 \hbar R_1^2 R_2^2}{3^3 \pi r^2 c^3} \frac{3^3 \pi c^3 \omega_2}{2^5 R_1^2 R_2} = \frac{\hbar \omega_2 R_2}{r^2} = \left(\frac{n_1 n_2 e^2}{r^2}\right)\left(\frac{n_2 \hbar \omega_2}{n_1 m_2 c^2}\right).$$
(13)

The expression for the force (13) is like the expression for the Coulomb force, $F_{12}(r) = n_1 n_2 e^2/r^2$, if

$$\frac{n_2 \hbar \omega_2}{n_1 m_2 c^2} = 1 \text{ or } \hbar \omega_2 = \frac{n_1 m_2 c^2}{n_2}.$$
(14)

Since for protons, electrons and muons, $n_p = n_e = n_\mu = 1$ or $n_1 = n_2 = 1$, then the condition for resonance is

$$\hbar \omega_i = m_i c^2, i = 1, 2.$$
(15)

The asymmetry of the expression of the force (13) is the result of the approximation of the integral $I_\Omega$ by the average $I_{\Omega av} = (I_{\Omega 1} + I_{\Omega 2})/2 \cong I_{\Omega 2}/2$ which was required by the necessity to obtain an analytical expression of the integrals.

### 3. The average scattering cross-section

At the microscopic level (relative to the CZPF background), the scattering cross-section of an electrically charged particle depends on the angular frequency $\omega$. In order to obtain the scattering cross-section independent of the angular frequency, we will average the expression of the cross-section given by Eq. (4). According to Appendix 3, the averaged scattering cross-section has the expression

$$\langle \sigma_s \rangle_\omega = \frac{8\pi^2 c R_i \omega_i^3}{\Omega^4}.$$
(16)

Assuming the energy density of the CZPF radiation background, $\int_0^\infty \rho(\omega)d\omega$, is infinite, then it is necessary to enter a limit for the angular frequency, i.e., $\Omega > \omega_i$. If this is the Thomson cross-section, $8\pi R_i^2/3$, then the value of the maximum angular frequency $\Omega$ is

$$\Omega(\omega_i) = \omega_i \left(\frac{3\pi c}{\omega_i R_i}\right)^{1/4} = \omega_i \left(\frac{3\pi m_i c^3}{\omega_i n_i^2 e^2}\right)^{1/4} = \omega_i \left(\frac{3\pi \hbar c}{n_i^2 e^2}\right)^{1/4} > \omega_i$$
(17)

and it depends on the natural angular frequency $\omega_i$.

This limit of the frequency must be also used to calculate numerically the integral of the interaction force between two oscillators. The difficulty that arises when calculating the integral is that this frequency is different for the two types of particles. And this occurs because the frequency depends on the natural frequency. One can solve this issue by improving the particle model. We assume to be necessary to consider the phenomenon of coupling of the oscillators in interaction. Our assumption is based on the analogy with the phenomenon that occurs at the interaction of several bubbles in a cluster [44]. Also, because the oscillators have an accelerated motion, they perceive the CZPF background as a Planckian radiation background with temperatures proportional to the average acceleration [26]. The relative average acceleration of the two oscillators depends on the parameters of the two interacting particles.

## 4. Conclusions

In the particle model proposed in this paper, the interaction of the oscillator with the homogeneous background of CZPF radiation generates, by absorption-emission (scattering), an isotropic and inhomogeneous background. Then this primary scattering generates a secondary radiation addressed to their reciprocal scattering and which can depict the electrostatic interaction.

The source of the energy of the interaction is the energy absorbed and scattered from the background of CZPF [45]. The electric charge is a measure of the capacity of fundamental particles to scatter the CZPF radiation background. This result does not surprise us, because the SED is based on the Maxwel's equations. And these equations also involve the Coulomb force of interaction of two charges.

In our approach of the electrically charged particle is for interest to highlight the interpretation of electric charge as the scattering capacity of the CZPF background. Analogous to the quantum model (in QED, particles change energy carried by photons), particles modelled in SED change energy carried by electromagnetic waves. This phenomenon has an analogous approach in the physics of the interaction between two oscillating bubbles in liquid [35-37, 46, 47]. The electrical acoustic charge is of two types because oscillators can oscillate in phase or phase opposition. Highlighting this property is only possible by modelling the interactions between the two oscillators in the SED formalism, without using the notion of interaction cross-section. The spin problem for the two-dimensional oscillator was solved in the paper [48]. The particle model proposed in the SED is also important for highlighting an attractive interaction when the oscillators absorb some of the scattered energy from the CZPF background. This phenomenon we will study in a forthcoming paper.

The proposed model also has shortcomings. The Zitterbewegung condition obtained in classical relativistic model of the electron [49] is $\hbar\omega_0 = 2mc^2$ and not the one used in our paper $\hbar\omega_0 = mc^2$. This discrepancy may exist because the motion of the oscillator is treated non-relativistically. Also, the model does not explain the nature of the elastic field that ensures the internal oscillation. A possible solution to this problem would be to model the charged particles as an oscillating vacuum bubble [50].

**Appendix 1. The Integral $I_\Omega$**

$$I_\Omega = \int_0^\Omega \frac{\omega^{11} d\omega}{\left[\left(\omega_0^2 - \omega^2\right)^2 + \left(2R_1/(3c)\right)^2 \omega^6\right]\left[\left(\omega_0^2 - \omega^2\right)^2 + \left(2R_2/(3c)\right)^2 \omega^6\right]} =$$

(A1.1)

$$\int_0^\Omega \frac{\omega^{11} d\omega}{\left[\left(\omega_0 - \omega\right)^2 \left(\omega_0 + \omega\right)^2 + \left(2R_1/(3c)\right)^2 \omega^6\right]\left[\left(\omega_0 - \omega\right)^2 \left(\omega_0 + \omega\right)^2 + \left(2R_2/(3c)\right)^2 \omega^6\right]} \cong$$

$$\int_{\omega_0}^{\omega_0 - \Omega} \frac{\omega_0^{11} (-dx)}{\left[4\omega_0^2 x^2 + \left(2R_1/(3c)\right)^2 \omega_0^6\right]\left[4\omega_0^2 x^2 + \left(2R_2/(3c)\right)^2 \omega_0^6\right]} =$$

$$\frac{(-\omega_0^7)}{2^4} \int_{\omega_0}^{\omega_0 - \Omega} \frac{dx}{\left[x^2 + \left(R_1 \omega_0^2/(3c)\right)^2\right]\left[x^2 + \left(R_2 \omega_0^2/(3c)\right)^2\right]} =$$

(A1.1)

$$\frac{(-\omega_0^7)}{2^4} \int_{\omega_0}^{\omega_0 - \Omega} \frac{dx}{x^4 + x^2 \left(\omega_0^2/(3c)\right)^2 \left(R_1^2 + R_2^2\right) + \left[R_1 R_2 \left(\omega_0^2/(3c)\right)^2\right]^2} = \frac{(-\omega_0^7)}{2^4} I_x,$$

$$\omega_0 - \omega = x, \Omega > \omega_0.$$

The $I_x$ integral is solved using formulas 2.161 and 2.103 of the book [41]

$$I_x = \int_{\omega_0}^{\omega_0 - \Omega} \frac{dx}{x^4 + x^2 \left(\omega_0^2/(3c)\right)^2 \left(R_1^2 + R_2^2\right) + \left[R_1 R_2 \left(\omega_0^2/(3c)\right)^2\right]^2} =$$

$$\frac{2}{\left(\omega_0^2/(3c)\right)^2 \left(R_1^2 - R_2^2\right)} \left[\int_{\omega_0}^{\omega_0 - \Omega} \frac{dx}{2x^2 + 2R_2^2 \left(\omega_0^2/(3c)\right)^2} - \int_{\omega_0}^{\omega_0 - \Omega} \frac{dx}{2x^2 + 2R_1^2 \left(\omega_0^2/(3c)\right)^2}\right] =$$

$$\frac{1}{\left(\omega_0^2/(3c)\right)^2 \left(R_1^2 - R_2^2\right)} \left[\int_{\omega_0}^{\omega_0 - \Omega} \frac{dx}{x^2 + \left(R_2 \omega_0^2/(3c)\right)^2} - \int_{\omega_0}^{\omega_0 - \Omega} \frac{dx}{x^2 + \left(R_1 \omega_0^2/(3c)\right)^2}\right] =,$$

$$\frac{1}{\left(\omega_0^2/(3c)\right)^2 \omega_0^2 \left(R_1^2 - R_2^2\right)} \left[\frac{3c}{R_2 \omega_0^2} \arctan \frac{3cx}{R_2 \omega_0^2}\bigg|_{\omega_0}^{\omega_0 - \Omega} - \frac{3c}{R_1 \omega_0^2} \arctan \frac{3cx}{R_1 \omega_0^2}\bigg|_{\omega_0}^{\omega_0 - \Omega}\right] =$$

$$\frac{(-3^3 \pi c^3)}{\omega_0^6 \left(R_1^2 - R_2^2\right)} \left(\frac{R_1 - R_2}{R_1 R_2}\right) = \frac{(-3^3 \pi c^3)}{\omega_0^6 R_1 R_2 \left(R_1 + R_2\right)}. \quad \frac{3c}{R_i \omega_0} \gg 1, \frac{3c\Omega}{R_i \omega_0^2} \gg 1, \Omega > \omega_0.$$

(A1.2)

$$h = \left(\omega_0^2/(3c)\right)^4 \left(R_1^2 + R_2^2\right)^2 - 4\left[R_1 R_2 \left(\omega_0^2/(3c)\right)^2\right]^2 = \left(\omega_0^2/(3c)\right)^4 \left(R_1^2 - R_2^2\right)^2 > 0,$$

$$\left(\omega_0^2/(3c)\right)^2 \left(R_1^2 + R_2^2\right) - \sqrt{h} = 2R_2^2 \left(\omega_0^2/(3c)\right)^2,$$

$$\left(\omega_0^2/(3c)\right)^2 \left(R_1^2 + R_2^2\right) + \sqrt{h} = 2R_1^2 \left(\omega_0^2/(3c)\right)^2, \text{if } R_1 \neq R_2.$$

(A1.3)

From the conditions $3c/(R_i\omega_0) \gg 1$ and $3c\Omega/(R_i\omega_0^2) \gg 1$, yields

$$\frac{3c}{R_i\omega_0} = \frac{3}{n_i^2}\frac{\hbar c}{e^2}\frac{m_i c^2}{\hbar\omega_0} \gg 1 \text{ sau } \frac{m_i c^2}{\hbar\omega_0} \gg \frac{n_i^2}{3}\frac{e^2}{\hbar c},$$

$$\frac{3c\Omega}{R_i\omega_0^2} = \frac{3}{n_i^2}\frac{\hbar c}{e^2}\frac{m_i c^2}{\hbar\omega_0}\frac{\Omega}{\omega_0} \gg 1 \text{ sau } \frac{m_i c^2}{\hbar\omega_0}\frac{\Omega}{\omega_0} \gg \frac{n_i^2}{3}\frac{e^2}{\hbar c}.$$

(A1.4)

## Appendix 2. The Integrals $I_{\Omega 1}$ and $I_{\Omega 2}$

$$I_{\Omega 1} = \int_0^\Omega \frac{\omega^{11} d\omega}{\left[(\omega_1^2 - \omega^2)^2 + (2R_1/(3c))^2 \omega^6\right]\left[(\omega_2^2 - \omega^2)^2 + (2R_2/(3c))^2 \omega^6\right]} =$$

$$\int_{\omega_1}^{\omega_1-\Omega} \frac{\omega_1^{11}(-dx)}{4\omega_1^6\left(x^2 + (R_1\omega_1^2/(3c))^2\right)\left[(\omega_2^2/\omega_1^2 - 1)^2 + (2R_2\omega_1/(3c))^2\right]} =$$

$$\frac{\omega_1^5}{4\left[(\omega_2^2/\omega_1^2 - 1)^2 + (2R_2\omega_1/(3c))^2\right]}\int_{\omega_1}^{\omega_1-\Omega} \frac{dx}{\left[x^2 + (R_1\omega_1^2/(3c))^2\right]} \cong$$

$$\frac{\omega_1^5}{4\left[(\omega_2^2/\omega_1^2 - 1)^2 + (2R_2\omega_1/(3c))^2\right]}\left[\frac{3c}{R_1\omega_1^2}\arctan\frac{3cx}{R_1\omega_1^2}\Big|_{\omega_1}^{\omega_1-\Omega}\right] \cong$$

$$\frac{(-\omega_1^5)}{4\left[(\omega_2^2/\omega_1^2 - 1)^2 + (2R_2\omega_1/(3c))^2\right]}\left(\frac{-3c\pi}{R_1\omega_1^2}\right) =$$

$$\frac{3\pi c\omega_1^3}{2^2\left[(\omega_2^2/\omega_1^2 - 1)^2 + (2R_2\omega_1/(3c))^2\right]} \cong \frac{3\pi c\omega_1^7}{2^2 \omega_2^4},$$

(A2.1)

$$\omega_1 - \omega = x, \Omega > \omega_2 > \omega_1, \frac{3c}{R_1\omega_1} \gg 1, \frac{3c\Omega}{R_1\omega_1^2} \gg 1, \omega_2^4 \gg (2R_2\omega_1/(3c))^2.$$

$$I_{\Omega 2} = \int_{\omega_2}^{\omega_2-\Omega} \frac{\omega_1^{11}(-dy)}{4\omega_2^6\left[y^2+\left(R_2\omega_2^2/(3c)\right)^2\right]\left[\left(\omega_1^2/\omega_2^2-1\right)^2+\left(2R_1\omega_2/(3c)\right)^2\right]} =$$

$$\frac{\omega_1^5}{2^2\left[\left(\omega_1^2/\omega_2^2-1\right)^2+\left(2R_1\omega_2/(3c)\right)^2\right]}\int_{\omega_2}^{\omega_2-\Omega}\frac{dy}{\left[y^2+\left(R_2\omega_2^2/(3c)\right)^2\right]} \cong$$

$$\frac{\omega_2^5}{2^2\left[\left(\omega_1^2/\omega_2^2-1\right)^2+\left(2R_1\omega_2/(3c)\right)^2\right]}\left[\frac{3c}{R_2\omega_2^2}\arctan\frac{3cy}{R_2\omega_2^2}\Big|_{\omega_2}^{\omega_2-\Omega}\right] \cong$$

$$\frac{(-\omega_2^5)}{2^2\left[\left(\omega_1^2/\omega_2^2-1\right)^2+\left(2R_1\omega_2/(3c)\right)^2\right]}\left(\frac{-3c\pi}{R_2\omega_2^2}\right) =$$

$$\frac{3\pi c\omega_2^3}{2^2\left[\left(\omega_1^2/\omega_2^2-1\right)^2+\left(2R_1\omega_2/(3c)\right)^2\right]} \cong \frac{3^3\pi c^3\omega_2}{2^4 R_1^2 R_2},$$
(A2.2)

$$\omega_2-\omega=y,\ \Omega>\omega_2>\omega_1,\ \frac{3c}{R_2\omega_2}\gg 1,\ \frac{3c\Omega}{R_2\omega_2^2}\gg 1,\ \left(2R_1\omega_2/(3c)\right)^2\gg 1.$$

The integrals, in the variables $x$ and $y$, are solved according to the formulas 2.103 of the book [41, p. 67]

**Appendix 3. The averaged scattering cross-section**

$$\langle\sigma_s\rangle_\omega = \frac{\int_0^\Omega \sigma_{si}(\omega)\rho(\omega)d\omega}{\int_0^\Omega \rho(\omega)d\omega} = \frac{\int_0^\Omega\left[\frac{8\pi R_i^2\omega^4}{3\left[\left(\omega_i^2-\omega^2\right)^2+\left(2R_i/(3c)\right)^2\omega^6\right]}\frac{\hbar\omega^3}{2\pi^2 c^3}\right]d\omega}{\int_0^\Omega \frac{\hbar\omega^3}{2\pi^2 c^3}d\omega},\ \Omega>\omega_i.\quad (A3.1)$$

To calculate the integral from the denominator, we use the property that the section has a maximum for the angular frequencies close to its natural angular frequency. In integral from the denominator we make the change of variable $\omega_i-\omega=x$ and it becomes

$$\langle\sigma_s\rangle_\omega = \frac{\int_0^\Omega \frac{8\pi R_i^2\omega^7 d\omega}{3\left[\left(\omega_i^2-\omega^2\right)^2+\left(2R_i/(3c)\right)^2\omega^6\right]}}{\int_0^\Omega \omega^3 d\omega} = \frac{\int_0^\Omega \frac{8\pi R_i^2\omega^7 d\omega}{3\left[\left(\omega_i^2-\omega^2\right)^2+\left(2R_i/(3c)\right)^2\omega^6\right]}}{\frac{\Omega^4}{4}} =$$

$$\frac{32\pi c R_i^2}{3\Omega^4}\int_0^\Omega \frac{\omega^7 d\omega}{\left[(\omega_i-\omega)^2(\omega_i+\omega)^2+\left(2R_i/(3c)\right)^2\omega^6\right]} \cong$$

$$\frac{32\pi c R_i^2\omega_i^5}{3\Omega^4}\int_{\omega_i}^{\omega_i-\Omega}\frac{(-dx)}{4\omega_i^2\left[x^2+\left(\omega_i^2 R_i/(3c)\right)^2\right]} \cong \frac{\left(-8\pi c R_i^2\omega_i^5\right)}{3\Omega^4}\left(\frac{-3\pi c}{\omega_i^2 R_i}\right) = \frac{8\pi^2 c R_i\omega_i^3}{\Omega^4}.$$
(A3.2)

The integral, in the variable $x$, are solved according to the formulas 2.103 of the book [41, p. 67].